\documentclass{article}
\usepackage{amsmath, amsfonts,graphicx, caption, subcaption,float}

\title{Zeta function for perturbed surfaces of revolution}
\author{Pedro Morales-Almazan}
\date{\texttt{pmorales@math.utexas.edu}}
\begin{document}
\maketitle

\begin{abstract}
In this paper we explore the Zeta function arising from a small perturbation on a surface of revolution and the effect of this on the functional determinant and in the change of the Casimir energy associated with this configuration.
\end{abstract}

\section{Introduction}
The Casimir effect arises in quantum field theory as a result of vacuum fluctuations of the electromagnetic field. It was first predicted by Hendrick Casimir in 1948 \cite{CasimirP48}. Since then, the Casimir effect has caught the attention of researchers in different fields such as physics, mathematics, and engineering. 

The Casimir energy results from considering all possible quantum fluctuations of the electromagnetic field in the vacuum. It is defined over all possible energy states of a quantum system, hence producing a divergent expression in most of the cases. The study of divergent expressions is customary in Quantum Field Theory, and it has lead to use and improve various regularization methods. Dimensional regularization, Green's function regularization, ultra-violet cut-off, and Zeta function regularization are methods commonly used in this context, among others. 

The use of a Zeta function was first introduced by Euler when working with series involving prime numbers \cite{Euler44, Euler60}. Later Riemann provided a complete analysis of his famous Zeta function and its analytic continuation \cite{Riemann92}. These ideas were then used by Littlewood and Hardy when studying problem involving ill defined quantities appearing in number theory \cite{Hardy16}.

Then, in the late 1970s, Dowker and Critchley \cite{Dowker76}, and Hawking \cite{Hawking77} suggested the use of zeta function regularizations in quantum physical problems. 

The nature of the Casimir effect makes it highly dependent on the geometry and the boundary conditions of the system, having both attractive and repulsive forces for different boundary conditions of the same configuration\cite{Kimball12, Kenneth02}. Hence a general theory is still elusive and it is necessary to analyze individual configurations. 

The original setting proposed by Casimir considered two parallel conducting plates in vacuum \cite{Casimir48}. After this, many other configurations had been studied with different boundary conditions. Here we analyze surfaces of revolution with Dirichlet boundary conditions. There has been other approaches for cylindrical and prism-type configurations\cite{Abalo10,Abalo12} and this paper intends to complement these works.

This paper is structured in five sections. The first section is an introduction to the topic and a brief description of the paper. In the second section we find the zeta function corresponding to the Laplacian of a surface of revolution embedded in $\mathbb{R}^3$. We find an integral representation for this and then we provide the analytic continuation to the left of the convergence region. In the third section we find the expressions for the functional determinant and the Casimir energy for any given profile function. In the fourth section we introduce a small perturbation to the profile function and analyze what is the effect on the change in the Casimir energy. In the fifth section we consider the case of a constant profile function, i.e. a finite cylinder, and find the perturbation in the change of the Casimir energy for this configuration.

\section{Zeta Function for the unperturbed case}
\subsection{Eigenvalue equation}
Consider the spectral zeta function associated with the Laplacian defined on the surface of revolution $M$ obtained by revolving the graph
of $0\leq f(x)\in C^2(a,b)$ around the $x$-axis. Here, the Laplacian is found by using the metric inherited from the Euclidean metric on $\mathbb{R}^3$ and we consider Dirichlet boundary conditions.

Following the idea described in \cite{Jeffres12}, we can find the spectral zeta function associated with the Laplacian by a contour integral and
then its analytic continuation by subtracting asymptotic terms provided by a WKB expansion on the solutions of the radial part of the associated
eigenvalue equation for the Laplacian.

Hence, using coordinates $(x,\theta)$ on $M$, by \cite{Jeffres12}, we have that the Laplacian can be written as
\begin{equation}
\Delta =\frac{-1}{1+f'^2}\left(\frac{\partial^2}{\partial x^2}+\left(\frac{f'}{f}-\frac{f'f''}{1+f'^2}\right)\frac{\partial}{\partial x}+\frac{1+f'^2}{f^2}\frac{\partial^2}{\partial \theta^2}\right)\,.
\end{equation}
Therefore the eigenvalue problem reads
\begin{equation}\label{eigeq}
\Delta\phi=\lambda^2\phi\,,
\end{equation}
with Dirichlet conditions at $x=a$ and $x=b$.

\subsection{Integral representation}
Using the separation of variables $\phi(x,\theta)=X(x)\Theta(\theta)$ we have that (\ref{eigeq}) can be written as
\begin{eqnarray}
X''+\left(\frac{f'}{f}-\frac{f'f''}{1+f'^2}\right)X'+(1+f'^2)\left(\lambda^2-\frac{k^2}{f^2}\right)X=0\label{xeq}\,,\\
\Theta''+k^2\Theta=0\label{teq},
\end{eqnarray}
with Dirichlet conditions $X(a)=X(b)$ and periodic boundary conditions on $\Theta$. By (\ref{teq}) we have that $k\in\mathbb{Z}$. The boundary value problem (\ref{xeq}) can be restated as the initial value problem
\begin{equation}\label{xeqinit}
X''+\left(\frac{f'}{f}-\frac{f'f''}{1+f'^2}\right)X'+(1+f'^2)\left(\lambda^2-\frac{k^2}{f^2}\right)X=0
\end{equation}
with the initial conditions $X_k(a;\lambda)=0$ and $X_k'(a;\lambda)=1$. From here we have that the eigenvalues $\lambda$ are the solutions for the equation
\begin{equation}
X_k(b;\lambda)=0\,.
\end{equation}

With this, we can write the zeta function associated with the Laplacian on $M$ as the contour integral
\begin{equation}\label{zetaint}
\zeta_\Delta(s)=\frac{1}{2\pi \imath}\sum_{k=\infty}^\infty\int_\gamma d\lambda\,\lambda^{-2s}\frac{d}{d\lambda}\log X_k(b;\lambda)\,,
\end{equation}
where $\gamma$ is a contour enclosing the eigenvalues $\lambda$. By Deforming the contour $\gamma$ into the imaginary axis, we can rewrite (\ref{zetaint}) as
\begin{eqnarray}\label{zetaint2}
\zeta_\Delta(s)=\frac{\sin(\pi s)}{\pi}\int_0^\infty d\lambda\,\lambda^{-2s}\frac{d}{d\lambda}\log X_0(b;\imath \lambda) \nonumber \\
+2\frac{\sin(\pi s)}{\pi}\sum_{k=1}^\infty\int_0^\infty d\lambda\,\lambda^{-2s}\frac{d}{d\lambda}\log X_k(b;\imath \lambda)\,.
\end{eqnarray}

\subsection{Analytic continuation}
In order to perform the analytic continuation of $(\ref{zetaint2})$ we have to separate the case of $k=0$ and $k\neq 0$.
\subsubsection{k=0 asymptotic expansion}\label{k0ass}

For $k=0$, we have that (\ref{xeqinit}) becomes
\begin{equation}\label{kzero}
X''+\left(\frac{f'}{f}-\frac{f'f''}{1+f'^2}\right)X'-\lambda^2(1+f'^2)X=0\,.
\end{equation}

In order to find the asymptotic expansion of $X_0(b;\imath \lambda)$, we use the WKB method by substituting
\begin{equation}\label{s0wkb}
S(x;\lambda)=\frac{\partial}{\partial x}\log X_0(x;\imath\lambda)
\end{equation}
into (\ref{kzero}). This gives
\begin{equation}
S'= (1+f'^2)\lambda^2-S^2-\left(\frac{f'}{f}-\frac{f'f''}{1+f'^2}\right)S\,.
\end{equation}
Suppose that $S$ has an asymptotic expansion for $\lambda\to\infty$ given by
\begin{equation}\label{s0asym}
S(x;\lambda)\sim\sum_{i=-1}^\infty s_i(x)\lambda^{-i}\,,
\end{equation}
then we have that $s_i$ can be found recursively by
\begin{equation}
s_{i+1}=\frac{1}{-2s_{-1}}\left(s_i'+\left(\frac{f'}{f}-\frac{f'f''}{1+f'^2}\right)s_{i}+\sum_{j=0}^i s_js_{i-j}+g_{i+1}\right),
\end{equation}
where
\begin{equation}
s_{-1}=\sqrt{1+f'^2}
\end{equation}
\begin{equation}
g_{-1}=1+f'^2,\quad g_i=0 \text{ for } i>-1\,.
\end{equation}


Therefore we can find the asymptotic expansion  as $\lambda\to\infty$ of $\log X_0(b;\imath\lambda)$ by substituting (\ref{s0asym}) into (\ref{s0wkb}),
\begin{equation}
\log X_0(b;\imath\lambda)\sim \log A^++\sum_{i=-1}^\infty \lambda^{-i}\int_a^b dt\, s_i(t)\,,
\end{equation}
where $A^+$ can be found from the initial conditions as
\begin{equation}
\log A^+=-\log(2\lambda s_{-1}(a))-\log\left(1+\sum_{j=1}^\infty\frac{s_{2j-1}(a)}{s_{-1}(a)}\lambda^{-2j}\right)\,.
\end{equation}

From this expression, we find the zeta function associated with $k=0$ to be
\begin{equation}
\zeta^0(s)=Z^0(s)+\sum_{i=-1}^{N-2}A_i^0(s)\,,
\end{equation}
where $Z^0(s)$ is given by
\begin{eqnarray}
Z^0(s)=\frac{\sin(\pi s)}{\pi}\int_0^1 d\lambda\,\lambda^{-2s}\frac{d}{d\lambda}\log X_0(b;\imath\lambda) \nonumber \\
+\frac{\sin(\pi s)}{\pi}\int_1^\infty d\lambda\,\lambda^{-2s}\frac{d}{d\lambda}\nonumber\\
\times\left(\log X_0(b;\imath\lambda)-\log A^+-\sum_{i=-1}^{N-2}\lambda^{-i}\int_a^b dt\, s_i(t)\right)\,,
\end{eqnarray}
and where the $A^0_i(s)$ are given by the contributions coming from $\lambda^{-i}$.









\subsubsection{$k\neq0$ asymptotic expansion}\label{knass}

In the case of $k\neq 0$ we perform a uniform asymptotic expansion in both $k$ and $\lambda$ by means of the substitution
\begin{equation}
\lambda=uk\,.
\end{equation}

With this, (\ref{xeqinit}) takes the form
\begin{equation}\label{knonzero}
X''+\left(\frac{f'}{f}-\frac{f'f''}{1+f'^2}\right)X'-\left(\frac{(1+u^2f^2)(1+f'^2)}{f^2}\right)k^2X=0
\end{equation}
with initial conditions $X_k(a;\imath\lambda)=0$ and $X_k'(a;\imath\lambda)=1$. Thus the eigenvalues are solutions to the equation
\begin{equation}
X_k(b,\imath\lambda)=0\,.
\end{equation}

To find the asymptotic expansion of $X_k(b;\imath\lambda)$ for both $k,\lambda\to\infty$, use WKB by considering
\begin{equation}
W(x,k)=\frac{\partial}{\partial x}X_k(x;\imath uk)\,.
\end{equation}
Therefore (\ref{knonzero}) reads
\begin{equation}\label{snasym}
W'=\left(\frac{(1+u^2f^2)(1+f'^2)}{f^2}\right)k^2-W^2-\left(\frac{f'}{f}-\frac{f'f''}{1+f'^2}\right)W\,.
\end{equation}

Consider the asymptotic expansion for $k\to\infty$,
\begin{equation}
W(x,k)\sim\sum_{i=-1}^\infty w_i(x)k^{-i}\,,
\end{equation}
then we have that $w_i(x)$ can be found recursively by (\ref{snasym}) as
\begin{equation}
w_{i+1}=-\frac{1}{2w_{-1}}\left(w_i'+\left(\frac{f'}{f}-\frac{f'f''}{1+f'^2}\right)w_i+\sum_{j=0}^iw_jw_{i-j}+h_{i+1}\right)\,,
\end{equation}
where
\begin{eqnarray}
w_{-1}=\sqrt{\frac{(1+u^2f^2)(1+f'^2)}{f^2}}\,,\nonumber\\
h_{-1}=\frac{(1+u^2f^2)(1+f'^2)}{f^2}\,,\quad h_i=0 \text{ for }i>-1\,.
\end{eqnarray}





Therefore, we can find the asymptotic expansion  as $k\to\infty$ of $\log X_k(b;\imath uk)$ as before,
\begin{equation}
\log X_k(b;\imath uk)\sim \log B^++\sum_{i=-1}^\infty k^{-i}\int_a^b dt\, w_i(t)\,,
\end{equation}
where $B^+$ can be found from the initial conditions as
\begin{equation}
\log B^+=-\log(2k w_{-1}(a))-\log\left(1+\sum_{j=1}^\infty\frac{w_{2j-1}(a)}{w_{-1}(a)}k^{-2j}\right)\,.
\end{equation}

From this expression, we find the zeta function associated with $k\neq0$ to be
\begin{equation}
\zeta^{\neq}(s)=Z^{\neq}(s)+\sum_{i=-1}^{N-2}A_i^{\neq}(s)\,,
\end{equation}
where $Z^{\neq}(s)$ is given by
\begin{eqnarray}
Z^{\neq}(s)=\frac{2\sin(\pi s)}{\pi}\sum_{k=1}^\infty\int_0^\infty d(uk)\,(uk)^{-2s}\frac{d}{d(uk)}\left(\log X_k(b;\imath uk)\right.\nonumber\\
\left.-\log B^+-\sum_{i=-1}^{N-2}k^{-i}\int_a^b dt\, w_i(t)\right)\,,
\end{eqnarray}
and the $A^{\neq}_i(s)$ are given by contribution coming from $k^{-i}$.

\section{Functional determinant and Casimir energy}

In order to find the functional determinant and the Casimir energy associated with this configuration, we need to extend the region of convergence of
the spectral zeta function in order to include the points $s=0$ and $s=-1/2$. This can be done by subtracting $N=3$ and $N=4$ terms respectively.

\subsection{Functional determinant}
To include $s=0$ in the convergence region for the integral representation of $\zeta_\Delta$, we need to subtract $N=3$ asymptotic terms. Thus We need to find $Z^0\,'(0), Z^{\neq}\,'(0)$, and $A^0_i\,'(0),A^{\neq}_i\,'(0)$ for $i=-1,0,1$.

From Section \ref{k0ass} we find the finite term and the asymptotic terms coming from the WKB expansion to be

\begin{equation}
Z^0\,'(0)=-\log X_0(b;0)-\log(2s_{-1}(a))+\sum_{i=-1}^1\int_a^bdt\, s_i(t)\,,
\end{equation}

\begin{equation}
A^0_{-1}\,'(0)=-\int_a^bdt\,\sqrt{1+f'^2}\,,
\end{equation}

\begin{equation}
A^0_0\,'(0)=0\,,
\end{equation}

\begin{equation}
A^0_1\,'(0)=-\int_a^bdt\, \left(\frac{f'^2}{8f^2(1+f'^2)^{1/2}}+\frac{f''}{4f(1+f'^2)^{3/2}}\right)\,.
\end{equation}

Similarly, we have that for the non-zero modes Section \ref{knass} gives us the finite and the asymptotic terms coming from the WKB expansion, 

\begin{eqnarray}
Z^{\neq}\,'(0)=-2\sum_{k=1}^\infty \left(\log X_k(b;0)+\left.\log(2kw_{-1}(a))\right|_{u=0}\right.\nonumber\\
\left.-\sum_{i=-1}^1k^{-i}\int_a^b dt\, \left.w_i(t)\right|_{u=0}\right)\,,
\end{eqnarray}

\begin{equation}
A^{\neq}_{-1}\,'(0)=\frac{1}{6}\int_a^bdt\, f^{-1}\sqrt{1+f'^2}\,,
\end{equation}

\begin{equation}
A^{\neq}_0\,'(0)=\frac{1}{2}\log(2\pi (f^2(a)+f^2(b)))\,,
\end{equation}

\begin{equation}
A^{\neq}_1\,'(0)=\frac{1}{6}\int_a^bdt\, \frac{f'^2}{\sqrt{1+f'^2}}+\frac{1}{2}\int_a^bdt\,\frac{ff''}{\sqrt{1+f'^2}}\,.
\end{equation}

\subsection{Casimir energy}

In order to find the Casimir energy, we need to calculate the residue $\zeta_\Delta$ at $s=-1/2$. For this, we evaluate each of the pieces at $s=-1/2$.

From Section \ref{k0ass} we have the following terms
\begin{eqnarray}
Z^0(-1/2)=-\frac{1}{\pi}\int_0^1 d\lambda\,\lambda\frac{d}{d\lambda}\log X_0(b;\imath\lambda)\nonumber\\
-\frac{1}{\pi}\int_1^\infty d\lambda\,\lambda \frac{d}{d\lambda}\left(\log X_0(b;\imath\lambda)-\log A^+-\sum_{i=-1}^{N-2}\lambda^{-i}\int_a^b dt\, s_i(t)\right)\,,
\end{eqnarray}

\begin{equation}
A^0_{-1}(-1/2)=\frac{1}{ 2\pi}\int_a^b dt \sqrt{1+f'(t)^2}\,,
\end{equation}

\begin{equation}
A^0_{0}(s)=-\frac{1}{\pi}\,,
\end{equation}

\begin{eqnarray}\label{res01_2}
\text{Res }A^0_{1}(-1/2)=\frac{1}{2\pi}\int_a^bdt\, \left(-\frac{f'(t)^2}{8f(t)^2\left(1+f'(t)^2\right)^{1/2}}\right.\nonumber\\
\left.+\frac{f''(t)}{4f(t)\left(1+f'(t)^2\right)^{3/2}}\right)\,,
\end{eqnarray}

\begin{eqnarray}
A^0_{2}(-1/2)=-\frac{1}{8\pi}\left(\frac{f'^2(a)+f'^4(a)-2f(a)f''(a)}{f^2(a)(1+f'^2(a))^2}\right.\nonumber\\
\left.+\frac{f'^2(b)+f'^4(b)-2f(b)f''(b)}{f^2(b)(1+f'^2(b))^2}\right)\,,
\end{eqnarray}

and for $k\neq0$, the following terms are found from Section \ref{knass},
\begin{eqnarray}
Z^{\neq}(-1/2)=-\frac{2}{\pi}\sum_{k=1}^\infty k\int_0^\infty du\,u\frac{d}{du}\left(\log X_k(b;\imath uk)\right.\nonumber\\
\left.-\log B^+-\sum_{i=-1}^{N-2}k^{-i}\int_a^b dt\, w_i(t)\right)\,,
\end{eqnarray}

\begin{equation}
A^{\neq}_{-1}(-1/2)=\frac{\zeta_R'(-2)}{\pi}\int_a^bdt\, f^{-2}\sqrt{1+f'^2}\,,
\end{equation}

\begin{equation}
A^{\neq}_{0}(-1/2)=\frac{1}{24}(f^{-1}(a)+f^{-1}(b))\,,
\end{equation}

\begin{equation}\label{resneq2_2}
\text{Res }A^{\neq}_{1}(-1/2)=\frac{1}{16\pi}\int_a^bdt\, \frac{f'^2}{f^2\sqrt{(1+f'^2)}}-\frac{1}{8\pi }\int_a^bdt\, \frac{f''}{f(1+f'^2)^{3/2}}\,,
\end{equation}

\begin{eqnarray}\label{resneq1_2}
\text{Res }A^{\neq}_{2}(-1/2)=-\frac{1}{256}\left(\frac{f^{-1}(a)f'^2(a)}{(1+f'^2(a))}+\frac{f^{-1}(b)f'^2(b)}{(1+f'^2(b))}\right)\nonumber\\
-\frac{1}{32}\left(\frac{f''(a)}{(1+f'^2(a))^2}+\frac{f''(b)}{(1+f'^2(b))^2}\right)\,,
\end{eqnarray}

\begin{equation}
\text{FP }A^{\neq}_{2}(-1/2)=\frac{1}{16}\int_a^bdt\, f^{-1}\frac{f'f''}{(1+f'^2)^{4}}\,.
\end{equation}

Notice that the residues $(\ref{res01_2})$ and $(\ref{resneq2_2})$ cancel each other, giving the residue of the zeta function at $s=-1/2$ to be
\begin{eqnarray}
\text{Res }\zeta(-1/2)=-\frac{1}{256}\left(\frac{f^{-1}(a)f'^2(a)}{(1+f'^2(a))}+\frac{f^{-1}(b)f'^2(b)}{(1+f'^2(b))}\right)\nonumber\\
-\frac{1}{32}\left(\frac{f''(a)}{(1+f'^2(a))^2}+\frac{f''(b)}{(1+f'^2(b))^2}\right)\,,
\end{eqnarray}
which depends only on boundary terms. In general this need not to be zero, wich would make the value of the energy for this configuration to be infinite. By considering the perturbed surface of revolution, we seek to find the change in the Casimir energy to be a finite number.

\section{Surface perturbation}
In this section we consider the effect of making a small perturbation on the profile function $f(x)$ by adding a localized bump $\epsilon g(x)$, where the support of $g(x)$ is a small interval around a point $c\in(a,b)$. Thus, we find the change in the Casimir energy on the manifold by considering the variational problem
\begin{equation}
\frac{d}{d\epsilon}\zeta_{\Delta\epsilon}(-1/2),
\end{equation}
where $\Delta_\epsilon$ is the Laplacian obtained by substituting $f(x)\to f(x)+\epsilon g(x)$ in the previous formalism.

In order to do this, we can find the series expansion for small $\epsilon$ for each of the zeta function terms up to $O(\epsilon^2)$.

\subsection{Asymptotic terms}
\subsubsection{ $k=0$ terms}

From the terms obtained in Section \ref{k0ass} , we make the series expansion of each individual term as a power series expansion in $\epsilon$, where we obtain

\begin{eqnarray}
A_{-1}^0(-1/2)=\frac{1}{2\pi}\int_a^bdt\,\sqrt{1+f'(t)^2}\nonumber\\
+\epsilon\frac{1}{2\pi}\int_a^bdt\,\frac{f'(t)}{\sqrt{1+f'(t)^2}}g'(t)+O(\epsilon^2)\,,
\end{eqnarray}

\begin{equation}
A_0^0(-1/2)=-\frac{1}{\pi}\,,
\end{equation}

\begin{eqnarray}
\text{Res }A^0_{1}(-1/2)=\frac{1}{2\pi}\int_a^bdt\, \left(-\frac{f'(t)^2}{8f(t)^2\left(1+f'(t)^2\right)^{1/2}}\right.\nonumber\\
\left.+\frac{f''(t)}{4f(t)\left(1+f'(t)^2\right)^{3/2}}\right)\nonumber\\
+\epsilon\frac{1}{2\pi}\int_a^bdt\,\frac{f'(t)^2}{4f(t)^3(1+f'(t)^2)^{1/2}}g(t)\nonumber\\
-\epsilon\frac{1}{2\pi}\int_a^bdt\,\frac{f'(t)(2+f'(t)^2)}{8f(t)^2(1+f'(t)^2)^{3/2}}g'(t)\nonumber\\
-\epsilon\frac{1}{2\pi}\int_a^bdt\,\frac{f''(t)}{4f(t)^2(1+f'(t)^2)^{3/2}}g(t)\nonumber\\
-\epsilon\frac{1}{2\pi}\int_a^bdt\,\frac{3f'(t)f''(t)}{4f(t)(1+f'(t)^2)^{5/2}}g'(t)\nonumber\\
+\epsilon\frac{1}{2\pi}\int_a^bdt\,\frac{1}{4f(t)(1+f'(t)^2)^{3/2}}g''(t)+O(\epsilon^2)\,,
\end{eqnarray}

\begin{eqnarray}
A^0_{2}(-1/2)=-\frac{1}{8\pi}\left(\frac{f'^2(a)+f'^4(a)-2f(a)f''(a)}{f^2(a)(1+f'^2(a))^2}\right.\nonumber\\
\left.+\frac{f'^2(b)+f'^4(b)-2f(b)f''(b)}{f^2(b)(1+f'^2(b))^2}\right)\,.
\end{eqnarray}

\subsubsection{$k\neq0$ terms}
Likewise, from Section \ref{knass} we obtain the power series expansion in $\epsilon$ for the asymptotic terms,
\begin{eqnarray}
A^{\neq}_{-1}(-1/2)=\frac{\zeta_R'(-2)}{\pi}\int_a^bdt\, f^{-2}\sqrt{1+f'^2}\nonumber\\
-\epsilon\frac{2\zeta_R'(-2)}{\pi}\int_a^bdt\,\frac{(1+f'(t)^2)^{1/2}}{f(t)^3}g(t)\nonumber\\
+\epsilon\frac{\zeta_R'(-2)}{\pi}\int_a^bdt\,\frac{f'(t)}{f(t)^2(1+f'(t)^2)^{1/2}}g'(t)+O(\epsilon^2)\,,
\end{eqnarray}

\begin{equation}
A^{\neq}_{0}(-1/2)=\frac{1}{24}(f^{-1}(a)+f^{-1}(b))\,,
\end{equation}

\begin{eqnarray}
\text{Res }A^{\neq}_{1}(-1/2)=-\frac{1}{2\pi}\int_a^bdt\, \left(-\frac{f'(t)^2}{8f(t)^2\left(1+f'(t)^2\right)^{1/2}}\right.\nonumber\\
\left.+\frac{f''(t)}{4f(t)\left(1+f'(t)^2\right)^{3/2}}\right)\nonumber\\
-\epsilon\frac{1}{2\pi}\int_a^bdt\,\frac{f'(t)^2}{4f(t)^3(1+f'(t)^2)^{1/2}}g(t)\nonumber\\
+\epsilon\frac{1}{2\pi}\int_a^bdt\,\frac{f'(t)(2+f'(t)^2)}{8f(t)^2(1+f'(t)^2)^{3/2}}g'(t)\nonumber\\
+\epsilon\frac{1}{2\pi}\int_a^bdt\,\frac{f''(t)}{4f(t)^2(1+f'(t)^2)^{3/2}}g(t)\nonumber\\
+\epsilon\frac{1}{2\pi}\int_a^bdt\,\frac{3f'(t)f''(t)}{4f(t)(1+f'(t)^2)^{5/2}}g'(t)\nonumber\\
-\epsilon\frac{1}{2\pi}\int_a^bdt\,\frac{1}{4f(t)(1+f'(t)^2)^{3/2}}g''(t)+O(\epsilon^2)\,,
\end{eqnarray}

\begin{eqnarray}
\text{Res }A^{\neq}_{2}(-1/2)=-\frac{1}{256}\left(\frac{f^{-1}(a)f'^2(a)}{(1+f'^2(a))}+\frac{f^{-1}(b)f'^2(b)}{(1+f'^2(b))}\right)\nonumber\\
-\frac{1}{32}\left(\frac{f''(a)}{(1+f'^2(a))^2}+\frac{f''(b)}{(1+f'^2(b))^2}\right)\,,
\end{eqnarray}

\begin{eqnarray}
\text{FP }A^{\neq}_{2}(-1/2)=\frac{1}{16}\int_a^bdt\, f^{-1}\frac{f'f''}{(1+f'^2)^{4}}\nonumber\\
-\frac{\epsilon}{16}\int_a^bdt \frac{f'(t)f''(t)}{f(t)^2(1+f'(t)^2)^4}g(t)+\frac{\epsilon}{16}\int_a^bdt \frac{f''(t)(1-7f'(t)^2)}{f(t)(1+f'(t)^2)^5}g'(t)\nonumber\\
+\frac{\epsilon}{16}\int_a^bdt \frac{f'(t)}{f(t)(1+f'(t)^2)^4}g''(t)+O(\epsilon^2)\,.
\end{eqnarray}

\subsection{Finite terms}
When considering the finite terms, we need to study the behavior of the eigenfunctions due to the perturbation as well as the impact of this perturbation on the WKB expansion coefficients $s_i$ and $w_i$.

\subsubsection{Perturbed eigenfunctions}

Here we analyze the change on the eigenfunctions $X_k(b; \xi)$ in the presence of the perturbation. In order to calculate the solutions of the perturbed equation $(\ref{xeqinit})$, we follow a similar approach as for the asymptotic terms.
We replace $f(x)\mapsto f(x)+\epsilon g(x)$ and perform an expansion in powers of $\epsilon$.
Therefore, $(\ref{xeqinit})$ becomes
\begin{eqnarray}\label{xeinitper}
\tilde{X}''+\left(\frac{f'+\epsilon g'}{f+\epsilon g}-\frac{\left(f'+\epsilon g'\right)\left(f''+\epsilon g''\right)}{1+\left(f'+\epsilon g'\right)^2}\right)\tilde{X}'\nonumber\\
+\left(1+\left(f'+\epsilon g'\right)^2\right)\left(\lambda^2-\frac{k^2}{(f+\epsilon g)^2}\right)\tilde{X}=0\,.
\end{eqnarray}
Thus, up to $O(\epsilon^2)$ terms, the perturbed equation with solution $\tilde{X}$ can be written as
\begin{equation}
F(\tilde{X}'',\tilde{X}',\tilde{X},x)+\epsilon G(\tilde{X}'',\tilde{X}',\tilde{X},x)=0,
\end{equation}
where $F$ is the original equation $(\ref{xeqinit})$ and $G$ is simply the coefficient of $\epsilon$ in the power series expansion.
Then, $G$ is computed to have the explicit form
\begin{eqnarray}\label{G}
G(\tilde{X}'',\tilde{X}',\tilde{X},x)=\left(\frac{fg'-gf'}{f^2}-\frac{f'3g''+f'g''-f'^2f''g'+f''g'}{(1+f'^2)^2}\right)\tilde{X}'\nonumber\\
+\left(\frac{2k^2(1+f'^2)g}{f^3}+2f'g'\left(\lambda^2-\frac{k^2}{f^2}\right)\right)\tilde{X}.
\end{eqnarray}
As it is customary with methods to solve perturbed differential equations\cite{Bush92, Shivamoggi03}, we can write the solution of $(\ref{xeinitper})$ as a combination of the unperturbed
problem and the first order perturbation
\begin{equation}
\tilde{X}=X+\epsilon \hat{X}.
\end{equation}
Following the general theory of perturbed differential equations\cite{Bush92,Shivamoggi03}, we have that $X$ and $\hat{X}$ satisfy the system of equations
\begin{eqnarray}
\label{Xori} F(X'',X',X,x)=0, \qquad X(a)=0, X'(a)=1\,,\\
\label{Xpart}F(\hat{X}'', \hat{X}',\hat{X},x)=-G(X'',X',X,x), \qquad \hat{X}(a)=0, \hat{X}(a)=0.
\end{eqnarray}
Notice that $(\ref{Xori})$ is just $(\ref{xeqinit})$, and that $(\ref{Xpart})$ is an inhomogeneous version of $(\ref{xeqinit})$ with vanishing initial conditions. Hence, we can find the solution to $(\ref{Xpart})$ by finding a particular solution for the inhomogeneous equation. 

Using variation of parameters, let $X_k^1$ and $X_k^2$ be a solutions set for $(\ref{xeqinit})$. Then we can write
\begin{equation}
\hat{X}_k(x)=v^1(x)X^1_k(x)+v^2(x)X^2_k(x),
\end{equation}
with
\begin{eqnarray}
v^1(x)= \int_{x_0}^x \frac{X^2_k(t) G(X'',X',X,t)}{W(X^1_k,X^2_k)(t)}dt\,,\\
v^2(x)= -\int_{x_0}^x \frac{X^1_k(t) G(X'',X',X,t)}{W(X^1_k,X^2_k)(t)}dt\,.
\end{eqnarray}

Therefore, we have that the perturbed solution can be written as
\begin{equation}
\tilde{X}_k(b;\xi)=X_k(b;\xi)+\epsilon \hat{X}_k(b;\xi)\,.
\end{equation}
With this, we find that $\log \tilde{X}_k(b;\xi)$ has an expansion
\begin{equation}
 \log \tilde{X}_k(b;\xi)=\log X_k(b;\xi)+\epsilon\frac{\hat{X}_k(b;\xi)}{X_k(b;\xi)}+O(\epsilon^2).
\end{equation}

\subsubsection{Perturbed WKB coefficients}
To obtain the perturbed WKB coefficients we follow the approach used in the previous section by replacing $f(x)$ by $f(x)+\epsilon g(x)$ and then finding a series expansion in $\epsilon$. We apply this to the WKB expansions obtained in Section \ref{k0ass} and Section \ref{knass} in order to obtain the terms up to $O(\epsilon^2)$.

From the WKB recursion equation in Section \ref{k0ass} and since $g^{(n)}(a)=0$ for all natural $n$, we have that $\log A^+$ remains the same after performing the perturbation. 

Also, from the WKB recursion in Section \ref{knass} and the vanishing derivatives of the perturbation function $g(x)$ at the borders, we have that $\log B^+$ also remains unchanged by the perturbation.

\subsection{Energy Perturbation}

Once we found the expansion of the perturbed zeta in terms of $\epsilon$, we can find the change in the energy by considering
\begin{equation}
\frac{d}{d\epsilon}\zeta_{\Delta_\epsilon}(-1/2)\,.
\end{equation}

Finding this change and taking the limit as $\epsilon\to0^+$ will give the instantaneous change in the Casimir energy produced by the perturbation. As the residue of the zeta function does not depend on $\epsilon$, we will have a well defined quantity.

\subsubsection{$k=0$ asymptotic terms}
By taking the derivative with respect to the perturbation parameter, we have that for $k=0$,
\begin{equation}
\left.\frac{d}{d\epsilon}A_{-1}^0(-1/2)\right|_{\epsilon=0}=\frac{1}{2\pi}\int_a^bdt\,\frac{f'(t)}{\sqrt{1+f'(t)^2}}g'(t)\,,
\end{equation}

\begin{equation}
\left.\frac{d}{d\epsilon}A_0^0(-1/2)\right|_{\epsilon=0}=0\,,
\end{equation}

\begin{eqnarray}
\left.\frac{d}{d\epsilon}\text{Res }A^0_{1}(-1/2)\right|_{\epsilon=0}=\frac{1}{2\pi}\int_a^bdt\,\frac{f'(t)^2}{4f(t)^3(1+f'(t)^2)^{1/2}}g(t)\nonumber\\
-\frac{1}{2\pi}\int_a^bdt\,\frac{f'(t)(2+f'(t)^2)}{8f(t)^2(1+f'(t)^2)^{3/2}}g'(t)\nonumber\\
-\frac{1}{2\pi}\int_a^bdt\,\frac{f''(t)}{4f(t)^2(1+f'(t)^2)^{3/2}}g(t)\nonumber\\
-\frac{1}{2\pi}\int_a^bdt\,\frac{3f'(t)f''(t)}{4f(t)(1+f'(t)^2)^{5/2}}g'(t)\nonumber\\
+\frac{1}{2\pi}\int_a^bdt\,\frac{1}{4f(t)(1+f'(t)^2)^{3/2}}g''(t)\,,
\end{eqnarray}

\begin{equation}
\left.\frac{d}{d\epsilon}A^0_{2}(-1/2)\right|_{\epsilon=0}=0\,.
\end{equation}

\subsubsection{$k\neq 0$ asymptotic terms}
Similarly, we have that the terms corresponding to $k\neq0$ give
\begin{eqnarray}
\left.\frac{d}{d\epsilon}A^{\neq}_{-1}(-1/2)\right|_{\epsilon=0}=-\frac{2\zeta_R'(-2)}{\pi}\int_a^bdt\,\frac{(1+f'(t)^2)^{1/2}}{f(t)^3}g(t)\nonumber\\
+\frac{\zeta_R'(-2)}{\pi}\int_a^bdt\,\frac{f'(t)}{f(t)^2(1+f'(t)^2)^{1/2}}g'(t)\,,
\end{eqnarray}

\begin{equation}
\left.\frac{d}{d\epsilon}A^{\neq}_{0}(-1/2)\right|_{\epsilon=0}=0\,,
\end{equation}

\begin{eqnarray}
\left.\frac{d}{d\epsilon}\text{Res }A^{\neq}_{1}(-1/2)\right|_{\epsilon=0}=-\frac{1}{2\pi}\int_a^bdt\,\frac{f'(t)^2}{4f(t)^3(1+f'(t)^2)^{1/2}}g(t)\nonumber\\
+\frac{1}{2\pi}\int_a^bdt\,\frac{f'(t)(2+f'(t)^2)}{8f(t)^2(1+f'(t)^2)^{3/2}}g'(t)\nonumber\\
+\frac{1}{2\pi}\int_a^bdt\,\frac{f''(t)}{4f(t)^2(1+f'(t)^2)^{3/2}}g(t)\nonumber\\
+\frac{1}{2\pi}\int_a^bdt\,\frac{3f'(t)f''(t)}{4f(t)(1+f'(t)^2)^{5/2}}g'(t)\nonumber\\
-\frac{1}{2\pi}\int_a^bdt\,\frac{1}{4f(t)(1+f'(t)^2)^{3/2}}g''(t)\,,
\end{eqnarray}

\begin{equation}
\left.\frac{d}{d\epsilon}\text{Res }A^{\neq}_{2}(-1/2)\right|_{\epsilon=0}=0\,,
\end{equation}

\begin{eqnarray}
\left.\frac{d}{d\epsilon}\text{FP }A^{\neq}_{2}(-1/2)\right|_{\epsilon=0}=-\frac{1}{16}\int_a^bdt \frac{f'(t)f''(t)}{f(t)^2(1+f'(t)^2)^4}g(t)\nonumber\\
+\frac{1}{16}\int_a^bdt \frac{f''(t)(1-7f'(t)^2)}{f(t)(1+f'(t)^2)^5}g'(t)+\frac{1}{16}\int_a^bdt \frac{f'(t)}{f(t)(1+f'(t)^2)^4}g''(t)\,.
\end{eqnarray}

\subsubsection{Finite terms}
Likewise, we have for the finite terms that
\begin{eqnarray}
\left.\frac{d}{d\epsilon}Z^0(-1/2)\right|_{\epsilon=0}=-\frac{1}{\pi}\int_0^1d\lambda\,\lambda\frac{d}{d\lambda} \frac{\hat{X}_0(b;\imath\lambda)}{X_0(b;\imath\lambda)}\nonumber\\
-\frac{1}{\pi}\int_1^\infty d\lambda\,\lambda\frac{d}{d\lambda}\left(\frac{\hat{X}_0(b;\imath\lambda)}{X_0(b;\imath\lambda)}-\sum_{i=-1}^2\lambda^{-i}\int_a^bdt\, \left.\frac{\partial}{\partial \epsilon}s_i(t)\right|_{\epsilon=0}\right)\,,
\end{eqnarray}
and
\begin{eqnarray}
\left.\frac{d}{d\epsilon}Z^{\neq}(-1/2)\right|_{\epsilon=0}\nonumber\\
=-\frac{2}{\pi}\sum_{k=1}^\infty k\int_0^\infty du\,u\frac{d}{du} \left(\frac{\hat{X}_k(b;\imath uk)}{X_k(b;\imath uk)}-\sum_{i=-1}^2k^{-i}\int_a^bdt\, \left.\frac{\partial}{\partial \epsilon}w_i(t)\right|_{\epsilon=0}\right)\,.
\end{eqnarray}

\subsubsection{Change on the Casimir Energy }
The residues from $A_1^0$ and $A_1^{\neq}$ cancel out, giving a finite value for the change in the Casimir energy
\begin{equation}
\Delta E=\left.\frac{\partial}{\partial\epsilon}\zeta_{\Delta_\epsilon}(-1/2)\right|_{\epsilon=0}\,.
\end{equation}

Therefore, the change in the Casimir energy is a well defined quantity given by
\begin{eqnarray}
\Delta E=\left.\frac{d}{d\epsilon}\zeta_{\Delta_\epsilon}(-1/2)\right|_{\epsilon=0}=-\frac{1}{2\pi}\int_a^bdt\,\frac{f''(t)}{\left(f'(t)^2+1\right)^{3/2}}g(t)\nonumber\\
 -\frac{\zeta_R'(-2)}{\pi}\int_a^b dt\, \frac{f(t) f''(t)+2 f'(t)^2+2}{f(t)^3 \left(f'(t)^2+1\right)^{3/2}}g(t)\nonumber\\
 +\frac{1}{16}\int_a^bdt\,  \frac{2 f'(t)^3\left(f'(t)^2+1\right)+f(t) f'(t) \left(5 f'(t)^2-3\right)f''(t)}{f(t)^3 \left(f'(t)^2+1\right)^5}g(t)\nonumber\\
 -\frac{1}{\pi}\int_0^1d\lambda\,\lambda\frac{d}{d\lambda} \frac{\hat{X}_0(b;\imath\lambda)}{X_0(b;\imath\lambda)}\nonumber\\
-\frac{1}{\pi}\int_1^\infty d\lambda\,\lambda\frac{d}{d\lambda}\left(\frac{\hat{X}_0(b;\imath\lambda)}{X_0(b;\imath\lambda)}-\sum_{i=-1}^2\lambda^{-i}\int_a^bdt\, \left.\frac{\partial}{\partial \epsilon}s_i(t)\right|_{\epsilon=0}\right)\nonumber\\
-\frac{2}{\pi}\sum_{k=1}^\infty k\int_0^\infty du\,u\frac{d}{du} \left(\frac{\hat{X}_k(b;\imath uk)}{X_k(b;\imath uk)}-\sum_{i=-1}^2k^{-i}\int_a^bdt\, \left.\frac{\partial}{\partial \epsilon}w_i(t)\right|_{\epsilon=0}\right)
\end{eqnarray}

\section{Constant profile function}
Here we consider the case of a finite cylinder using the formalism we developed. If we consider the cylinder given by $f(x)=\alpha$, $\alpha\in\mathbb{R}^+$, then we find that all the perturbed asymptotic terms vanish except
\begin{equation}
\left.\frac{d}{d\epsilon}A^{\neq}_{-1}(-1/2)\right|_{\epsilon=0}=-\frac{2\zeta_R'(-2)}{\pi \alpha^3}\int_a^bdt\,g(t)\,.
\end{equation}

Likewise we have that 
\begin{equation}
 \int_a^bdt\,\frac{\partial}{\partial\epsilon}s_i(t)=0\,,
\end{equation}
for $i=-1,0,1,2$. For $w_i$ we have that
\begin{equation}
 \int_a^bdt\,\frac{\partial}{\partial\epsilon}w_{-1}(t)=\frac{-1}{\alpha^2\sqrt{1+u^2\alpha^2}}\int_a^bdt\, g(t)
\end{equation}
and 
\begin{equation}
 \int_a^bdt\,\frac{\partial}{\partial\epsilon}w_i(t)=0
\end{equation}
for $i=0,1,2$. Therefore the change in the Casimir energy is given by
\begin{eqnarray}\label{casimircyl}
 \Delta E=-\frac{2\zeta_R'(-2)}{\pi \alpha^3}\int_a^bdt\,g(t)-\frac{1}{\pi}\int_0^\infty d\lambda\,\lambda\frac{d}{d\lambda}\frac{\hat{X}_0(b;\imath\lambda)}{X_0(b;\imath\lambda)}\nonumber\\
 -\frac{2}{\pi}\sum_{k=1}^\infty k\int_0^\infty du\,u\frac{d}{du}\left(\frac{\hat{X}_k(b;\imath uk)}{X_k(b;\imath uk)}+\frac{k}{\alpha^2\sqrt{1+u^2\alpha^2}}\int_a^bdt\, g(t)\right)\,.
\end{eqnarray}

For both $k=0$ and $k\neq0$ we can explicitly find the solutions of $(\ref{eigeq})$. In the case of $k=0$, we have that the solutions are given by 
\begin{equation}
 X_0(b,\imath\lambda)= \frac{\sinh ((b - a)\lambda)}{\lambda}\,,
\end{equation}
with a solution set given by
\begin{equation}
 X^1_0(x;\imath\lambda)=e^{x\lambda}\,,\qquad X^2_0(x;\imath\lambda)= e^{-x\lambda}\,,
\end{equation}
which have a Wronskian of
\begin{equation}
 W(X^1_0,X^2_0)(x;\imath\lambda)=-2\lambda\,.
\end{equation}
Here, we have that
\begin{equation}
 G(X_0'',X_0',X_0,x)=\frac{g'(x)}{\alpha}X_0'(x;\imath\lambda)\,,
\end{equation}
which gives
\begin{equation}
 v^1(x)=-\frac{1}{2\alpha\lambda}\int_{x_0}^x X^2_0(t)X_0'(t)g'(t)dt
\end{equation}
and
\begin{equation}
 v^2(x)=\frac{1}{2\alpha\lambda}\int_{x_0}^x X^1_0(t)X_0'(t)g'(t)dt\,.
\end{equation}
Applying integration by parts we get
\begin{equation}
 v^1(b)=\frac{1}{2\alpha\lambda}\int_a^b dt\, (X_0^2X_0')'(t)g(t)\,,
\end{equation}
\begin{equation}
 v^2(b)=-\frac{1}{2\alpha\lambda}\int_a^b dt\, (X_0^1X_0')'(t)g(t)\,,
\end{equation}
which in this case gives
\begin{equation}
 v^1(b)=-\frac{1}{2\alpha}\int_a^b dt\, e^{(a-2 t) \lambda } g(t)\,,
\end{equation}
\begin{equation}
 v^2(b)=-\frac{1}{2\alpha}\int_a^b dt\, e^{-(a-2 t) \lambda }g(t)\,.
\end{equation}
Therefore, we have that the first order perturbation $\hat{X}_0$ is given by
\begin{equation}
 \hat{X}_0(b;\imath\lambda)=-\frac{1}{\alpha}\int_a^bdt\, \text{cosh}((a+b-2 t) \lambda )g(t)
\end{equation}
and therefore
\begin{equation}\label{force0}
\frac{\hat{X}_0(b;\imath\lambda)}{X_0(b;\imath\lambda)}=-\frac{\lambda}{\alpha}\int_a^bdt\,  \text{cosh}((a+b-2 t) \lambda ) \text{csch}((b-a) \lambda )g(t)\,.
\end{equation}
Likewise, for $k\neq 0$ we find the explicit solutions given by
\begin{equation}
 X_k(x;\imath uk)= -\frac{\alpha  \sinh\left(\frac{k (a-t) \sqrt{1+u^2 \alpha ^2}}{\alpha }\right)}{k \sqrt{1+u^2 \alpha ^2}}
\end{equation}
and the independent set of solutions, 
\begin{equation}
 X^1_k(x;\imath uk)=e^{\frac{k t \sqrt{1+u^2 \alpha ^2}}{\alpha }}\,,
\end{equation}
\begin{equation}
 X^2_k(x;\imath uk)=e^{-\frac{k t \sqrt{1+u^2 \alpha ^2}}{\alpha }}\,,
\end{equation}
which have a Wronskian of
\begin{equation}
 W(X^1_k,X^2_k)(x)=-\frac{2 k \sqrt{1+u^2 \alpha ^2}}{\alpha }\,.
\end{equation}

In this case we find that 
\begin{equation}
 G(X''_k,X'_k,X_k,x)=\left(\frac{g'(x)}{\alpha}\right)X'_k+\left(\frac{2k^2g(x)}{\alpha^3}\right)X_k\,,
\end{equation}
\begin{equation}
 v^1(b)=-\frac{e^{-\frac{a k \sqrt{1+u^2 \alpha ^2}}{\alpha }}}{2 \left(\alpha +u^2 \alpha ^3\right)}\int_a^b dt\,  \left(1+e^{\frac{2 k (a-t) \sqrt{1+u^2 \alpha ^2}}{\alpha }} u^2 \alpha ^2\right)g(t)\,,
\end{equation}

\begin{equation}
 v^2(b)=-\frac{e^{\frac{a k \sqrt{1+u^2 \alpha ^2}}{\alpha }}}{2 \left(\alpha +u^2 \alpha ^3\right)}\int_a^b dt\,  \left(1+e^{\frac{2 k (-a+t) \sqrt{1+u^2 \alpha ^2}}{\alpha }} u^2 \alpha ^2\right)g(t)\,,
\end{equation}
and therefore a particular solution given by the variation of parameters is
\begin{eqnarray}
 \hat{X}_k(b;\imath uk)=-\frac{1}{\alpha +u^2 \alpha ^3}\int_a^bdt\, \left(\text{cosh}\left(\frac{(-a+b) k \sqrt{1+u^2 \alpha ^2}}{\alpha }\right)\right.\nonumber\\
 \left.+u^2 \alpha ^2 \text{cosh}\left(\frac{k (a+b-2 t) \sqrt{1+u^2 \alpha ^2}}{\alpha }\right)\right)g(t)\,,
\end{eqnarray}
\begin{eqnarray}\label{forcek}
 \frac{\hat{X}_k(b;\imath uk)}{X_k(b;\imath uk)}=-\frac{k\text{csch}\left(\frac{k(b-a) \sqrt{1+u^2 \alpha ^2}}{\alpha }\right)}{\alpha ^2 \sqrt{1+u^2 \alpha ^2}}\int_a^bdt\, \left(\text{cosh}\left(\frac{k(b-a)  \sqrt{1+u^2 \alpha ^2}}{\alpha }\right)\right.\nonumber\\
 \left.+u^2 \alpha ^2 \text{cosh}\left(\frac{k (a+b-2 t) \sqrt{1+u^2 \alpha ^2}}{\alpha }\right)\right) g(t)\,.
\end{eqnarray}

\subsection{Numerical Approximation}
Numerical methods can be used in order to better understand the behavior of the change in energy for the constant profile function setup. 

The expression for the change in the Casimir energy $(\ref{casimircyl})$ is made of three expression, that is, an integral of $g$, a double integral related to $\hat{X}_0/X_0$ ,and a series involving a double integral of $\hat{X}_k/X_k$.

The finite integrals over $[a,b]$ can be approximated by adaptive quadrature numerical methods up to any prescribed accuracy of absolute and relative errors. 

For the infinite integrals, we use integration by parts to remove the inner derivatives, so we have 
\begin{equation}\label{par1}
\int_0^\infty d\lambda\, \lambda \frac{d}{d\lambda}\frac{\hat{X}_0(b;\imath\lambda)}{X_0(b;\imath\lambda)}=\left.\lambda\frac{\hat{X_0}(b;\imath\lambda)}{X_0(b;\imath\lambda)}\right|_0^\infty-\int_0^\infty d\lambda\frac{\hat{X_0}(b;\imath\lambda)}{X_0(b;\imath\lambda)}\,,
\end{equation}
and 
\begin{eqnarray}\label{par2}
\int_0^\infty du\, u \frac{d}{du}\left(\frac{\hat{X}_k(b;\imath u k)}{X_k(b;\imath u k)}+\frac{k}{\alpha^2\sqrt{1+u^2\alpha^2}}\int_a^bdt\,g(t)\right)\nonumber\\
=\left.u\left(\frac{\hat{X_k}(b;\imath uk)}{X_k(b;\imath uk)}+\frac{k}{\alpha^2\sqrt{1+u^2\alpha^2}}\int_a^bdt\,g(t)\right)\right|_0^\infty\nonumber\\
-\int_0^\infty du\,\frac{\hat{X_k}(b;\imath uk)}{X_k(b;\imath uk)}+\frac{k}{\alpha^2\sqrt{1+u^2\alpha^2}}\int_a^bdt\,g(t)\,.
\end{eqnarray}

For both $(\ref{par1})$ and $(\ref{par2})$ we have that the boundary contribution vanish, hence giving 
\begin{equation}\label{term2}
\int_0^\infty d\lambda\, \lambda \frac{d}{d\lambda}\frac{\hat{X}_0(b;\imath\lambda)}{X_0(b;\imath\lambda)}=-\int_0^\infty d\lambda\frac{\hat{X_0}(b;\imath\lambda)}{X_0(b;\imath\lambda)}\,,
\end{equation}
and 
\begin{eqnarray}\label{term3}
\int_0^\infty du\, u \frac{d}{du}\left(\frac{\hat{X}_k(b;\imath u k)}{X_k(b;\imath u k)}+\frac{k}{\alpha^2\sqrt{1+u^2\alpha^2}}\int_a^bdt\,g(t)\right)\nonumber\\
=-\int_0^\infty du\,\frac{\hat{X_k}(b;\imath uk)}{X_k(b;\imath uk)}+\frac{k}{\alpha^2\sqrt{1+u^2\alpha^2}}\int_a^bdt\,g(t)\,.
\end{eqnarray}

Since the improper integrals converge, in order to treat them, we can change the unbounded domain of integration to a bounded one by performing a change of variables,
\begin{equation}
\int_0^\infty du\, f(u)=\int_0^1dt\, \frac{1}{(1-t)^2}f\left(\frac{t}{1-t}\right)\,.
\end{equation}
Hence, with these expressions, it is possible to use regular adaptive quadrature numerical methods up to any precision in order to calculate the improper integrals. 

Lastly, in order to consider the series in $k$ for the third term in $(\ref{casimircyl})$, we consider an approximation by including $N$ terms in the series such that the error given by the tails estimate,
\begin{equation}
R_N=\left| \int_N^\infty d\kappa\, t\int_0^\infty du\,\frac{\hat{X_\kappa}(b;\imath u\kappa)}{X_\kappa(b;\imath u\kappa)}+\frac{\kappa}{\alpha^2\sqrt{1+u^2\alpha^2}}\int_a^bdt\,g(t)\right|\,,
\end{equation} 
is smaller than any prescribed error. 

\subsection{Gaussian perturbation}
For this, we consider Gaussian perturbations to the profile function centered at $x=c$ of witdh $2\delta$, given by
\begin{equation}
g_\delta(x,c)=\chi(I)\exp\left(-\left(\frac{(x-c)}{(x-c)^2-\delta^2}\right)^2\right)\,,
\end{equation}
where $\chi(I)$ is the characteristic function of the interval $I=(c-\delta,c+\delta)$.

\begin{figure}[H]
\centering
\begin{subfigure}[H]{0.45\textwidth}
\includegraphics[width=\textwidth]{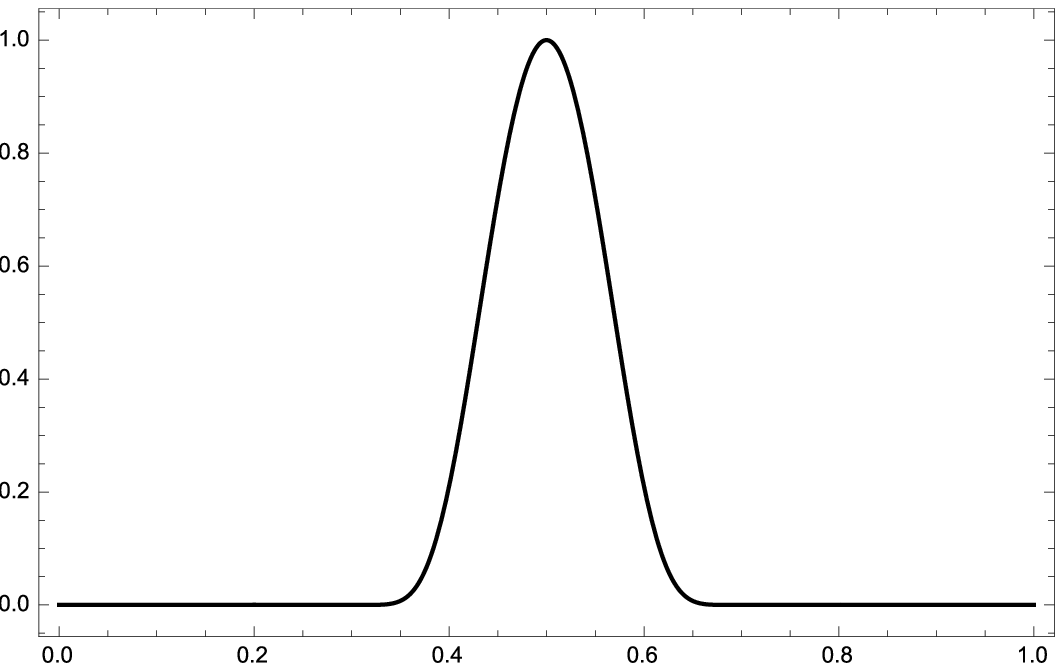}
 \label{figure:gep03}
 \caption{$\delta=0.3$}
\end{subfigure}
\begin{subfigure}[H]{0.45\textwidth}
\includegraphics[width=\textwidth]{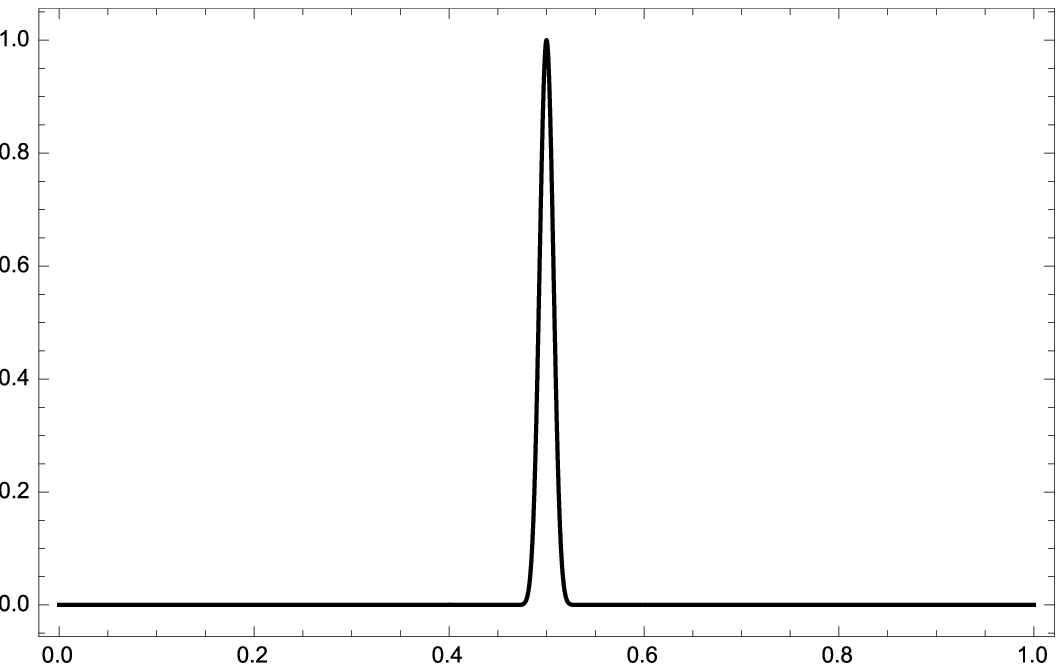}
 \label{figure:gep01}
 \caption{$\delta=0.1$}
\end{subfigure}
\caption{Gaussian potentials for different values of $\delta$}
\label{gaussian}
\end{figure}

\begin{figure}[H]
\centering
\begin{subfigure}[H]{0.45\textwidth}
\includegraphics[width=\textwidth]{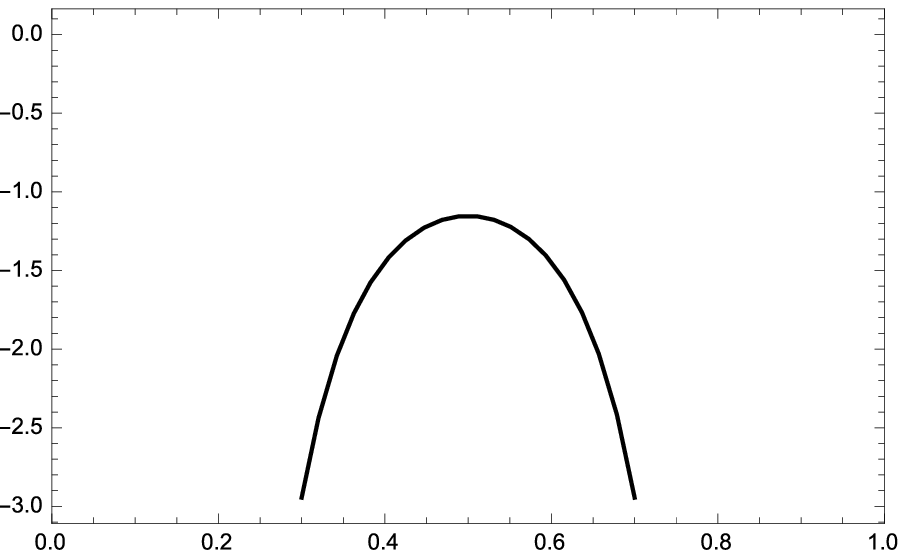}
 \label{figure:cep03}
 \caption{$\delta=0.3$}
\end{subfigure}
\begin{subfigure}[H]{0.45\textwidth}
\includegraphics[width=\textwidth]{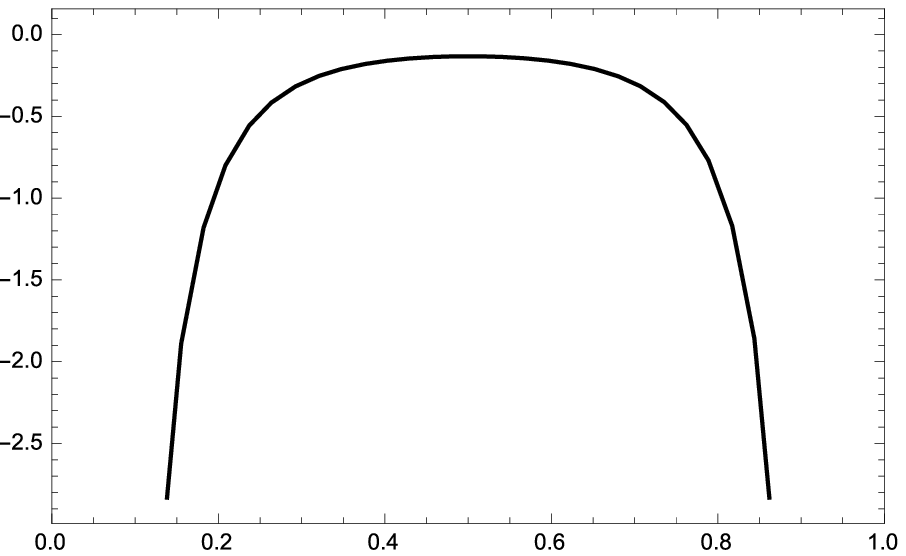}
 \label{figure:cep01}
 \caption{$\delta=0.1$}
\end{subfigure}
\caption{Change in the Casimir energy over $[0,1]$ for different values of $\delta$}
\label{casfigs}
\end{figure}

For the numerical analysis, we set $f(x)=\alpha=1$, and vary the length of the interval and the width of the perturbation. For an absolute error of $1\times 10^{-6}$, we have that the graph of the change in the Casimir energy with respect to the perturbation position $c$ is given in  Figure \ref{casfigs}.

We have then that the change in the Casimir energy is negative everywhere, having bigger absolute change near the edges of the interval. 

\begin{figure}[H]
\centering
\begin{subfigure}[H]{0.45\textwidth}
\includegraphics[width=\textwidth]{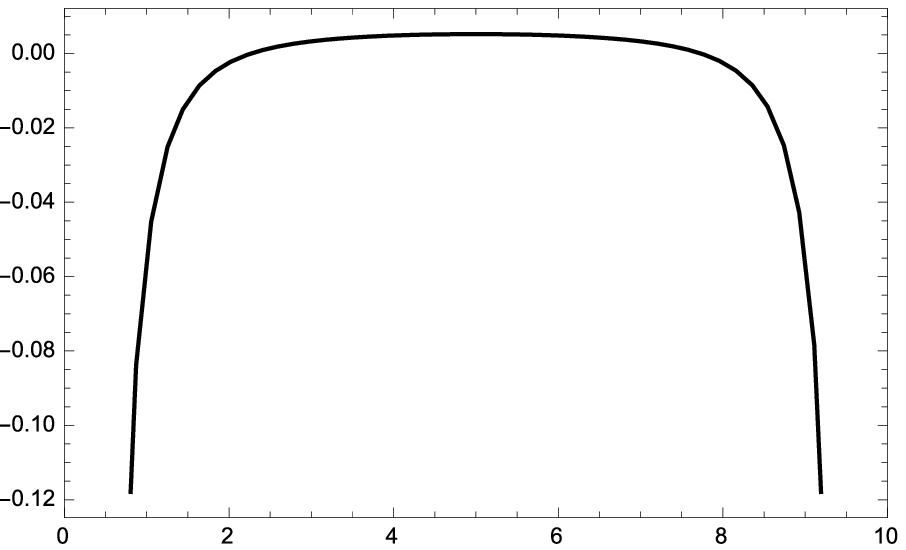}
 \label{figure:clep03}
 \caption{$\delta=0.3$}
\end{subfigure}
\begin{subfigure}[H]{0.45\textwidth}
\includegraphics[width=\textwidth]{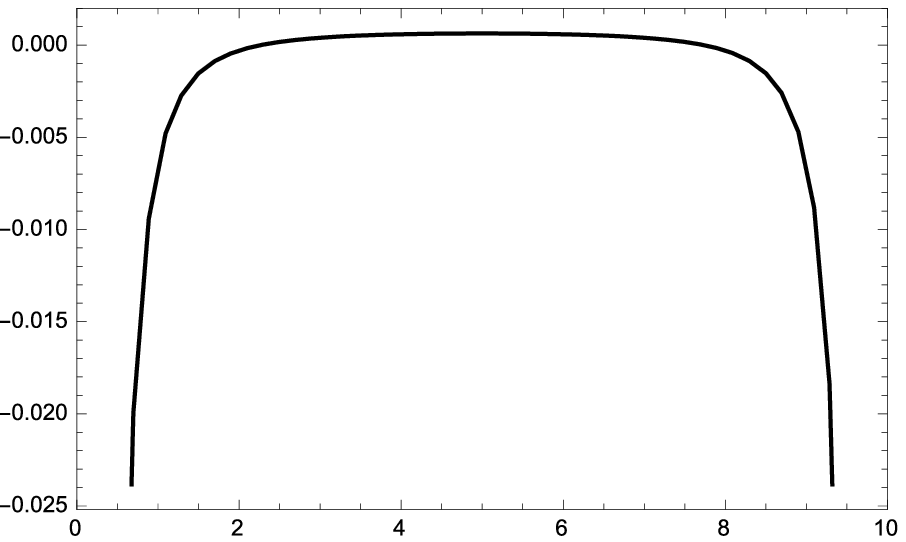}
 \label{figure:clep01}
 \caption{$\delta=0.1$}
 \end{subfigure}
\caption{Change in the Casimir energy over $[0,10]$ for different values of $\delta$}
\label{caslfigs}
\end{figure}

If the interval $[a,b]$ gets bigger, increasing the proportion between the radius $\alpha$ and the length of the interval, we have a different behavior in the change of the Casimir Energy. In Figure \ref{caslfigs} we show the numerical analysis made for the interval $[0,10]$ fixing the other parameters parameters.

Notice that there is a section in  the middle of the interval in Figure \ref{caslfigs} for which deformations made there will result in an approximately zero change in the Casimir energy. This means that the Casimir effect feels the edges and their proximity and the relation between the dimensions of the surface.

\begin{figure}[H]
\centering
\begin{subfigure}[H]{0.45\textwidth}
\includegraphics[width=\textwidth]{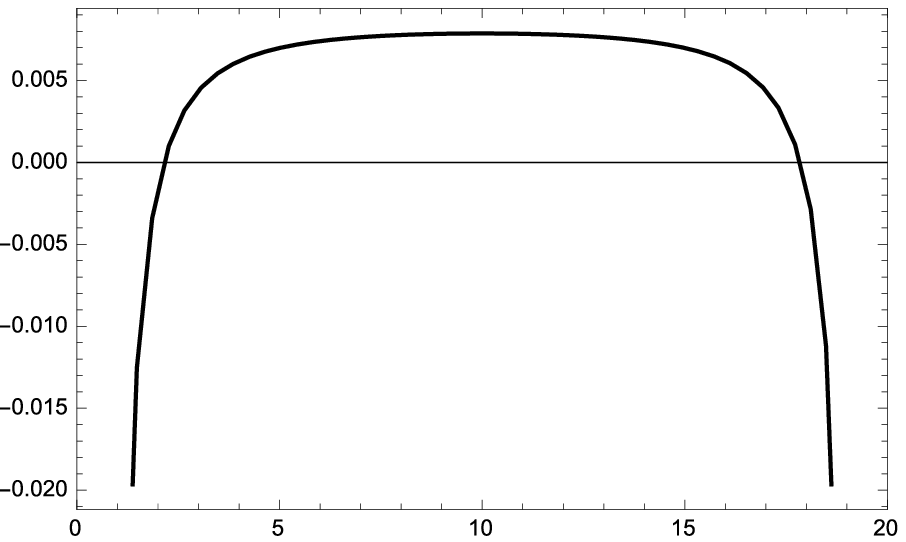}
 \label{figure:cl20ep03}
 \caption{$\delta=0.3$}
\end{subfigure}
\begin{subfigure}[H]{0.45\textwidth}
\includegraphics[width=\textwidth]{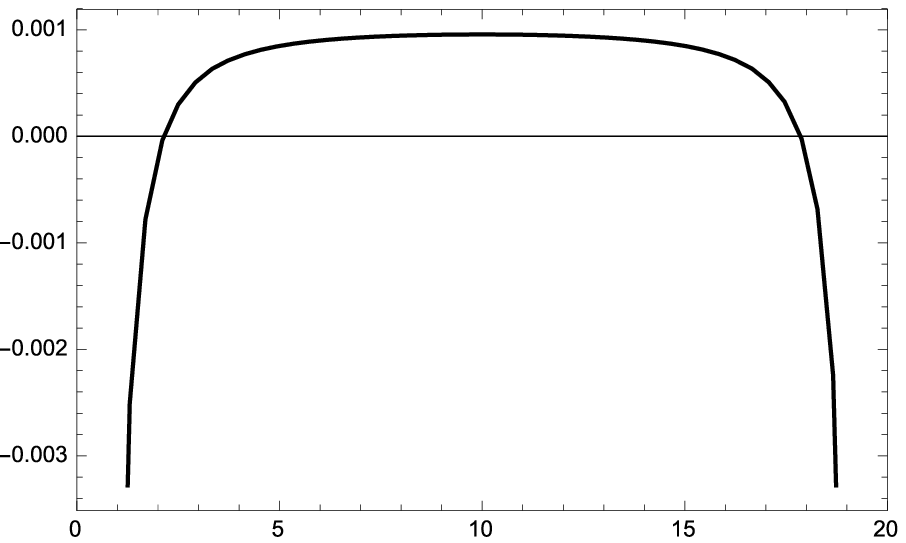}
 \label{figure:cl20ep01}
 \caption{$\delta=0.1$}
\end{subfigure}
\caption{Change in the Casimir energy over $[0,20]$ for different $\epsilon$}
\label{casl20figs}
\end{figure}

In Figure \ref{casl20figs} we have the graph of the change for the energy for the interval $[0,20]$. Here we can see that it changes from negative to positive. This behavior mainly occurs due to the proportions in the length of the interval and the radius $\alpha$ of the cylinder. It appears to be that the Casimir energy feels how close is the edge. Near the middle, where the edges are now farther away, the energy is starting to behave like the case of an infinite cylinder.
 
\begin{figure}[H]
\centering
\begin{subfigure}[H]{0.45\textwidth}
\includegraphics[width=\textwidth]{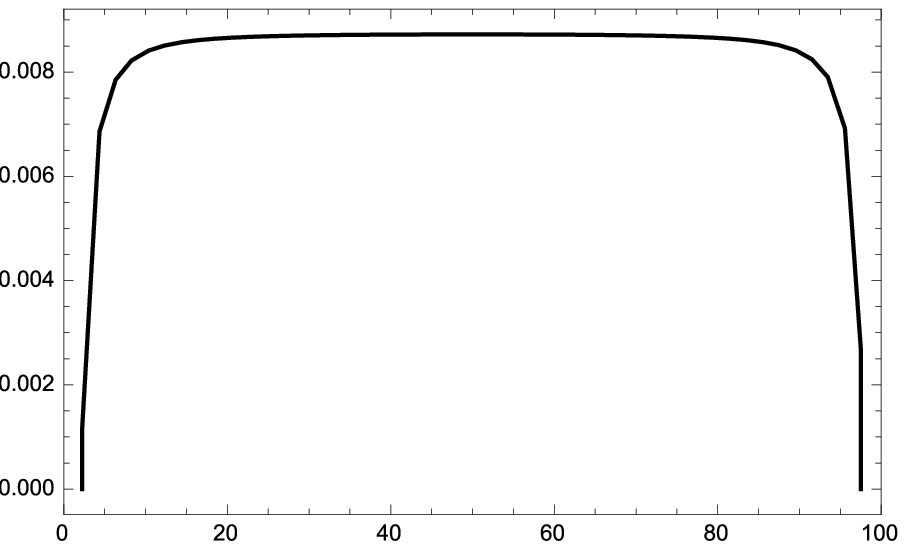}
 \label{figure:cl100ep03}
 \caption{$\delta=0.3$}
\end{subfigure}
\begin{subfigure}[H]{0.45\textwidth}
\includegraphics[width=\textwidth]{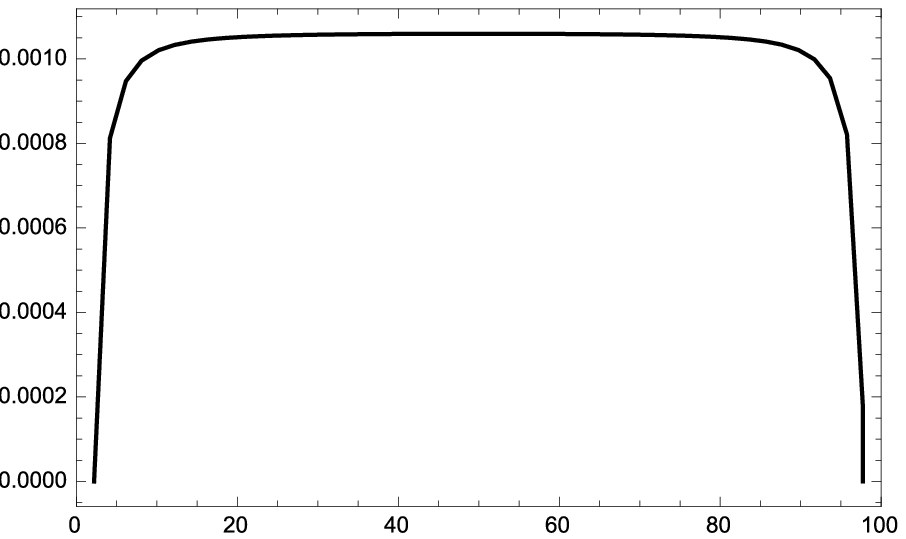}
 \label{figure:cl100ep01}
 \caption{$\delta=0.1$}
\end{subfigure}
\caption{Change in the Casimir energy over $[0,100]$ for different values of $\delta$}
\label{casl100figs}
\end{figure}

When increasing the length of the interval to $[0,100]$ in Figure \ref{casl100figs}, we have that the change in the energy is positive throughout the majority of the interval. When the perturbation is far away from the edges, then the system behaves like a similar system with no edges. In this case, it behaves like an infinite cylinder, for which the change in the Casimir energy tends to a positive value as $a,b\to\infty$, depending on how $|a|,|b|$ increase. This agrees with the results obtained in \cite{Gosdzinsky98}.

Notice that the integrands of $(\ref{force0})$ and $(\ref{forcek})$, when regarded as functions of $t$, are symmetric with respect to the midpoint of the surface, $t=\frac{a+b}{2}$. Therefore, after taking their derivatives with respect to $\lambda$ and $u$ respectively, we still have symmetric functions of $t$ around the midpoint. This means that the integrands in $(\ref{casimircyl})$, viewed as functions of $t$, are symmetric around the midpoint. Therefore, the change in the Casimir energy $(\ref{casimircyl})$ depends symmetrically on $c$ with respect to the midpoint $t=\frac{a+b}{2}$, and we can expect to have a maximum change in the Casimir energy at the midpoint of the interval, since this is the point that is the farthest away from the edges.  

\subsection{Mixed Gaussian perturbation}
\begin{figure}[H]
\centering
\includegraphics[width=0.6\textwidth]{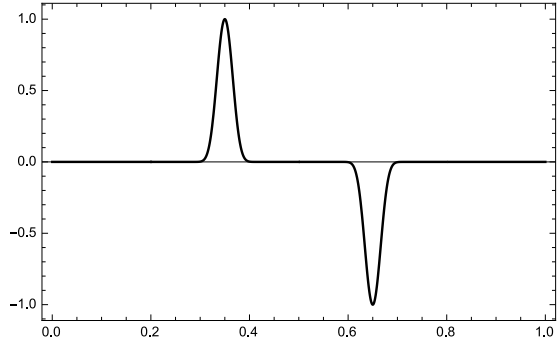}
\caption{Mixed Gaussian perturbation for $\delta=0.3$ }
\label{gausmix}
\end{figure}

We have a different behavior for a perturbation that is both positive and negative like in Figure \ref{gausmix}. Here we use a mixed Gaussian perturbation,

\begin{equation}
g_\delta(x,c) = 
\begin{cases}
\exp\left(-\left(\frac{(x-(c-\delta/2))}{(x-(c-\delta/2))^2-(\delta/2)^2}\right)^2\right)  &  c-\delta<x<c \\  
-\exp\left(-\left(\frac{(x-(c+\delta/2))}{(x-(c+\delta/2))^2-(\delta/2)^2}\right)^2\right)  &  c<x<c+\delta \\
0 & \mbox{otherwise.}
\end{cases}
\end{equation}

The first part is a positive Gaussian bump and the second part is a negative Gaussian bump.

\begin{figure}[H]
\centering
\includegraphics[width=0.6\textwidth]{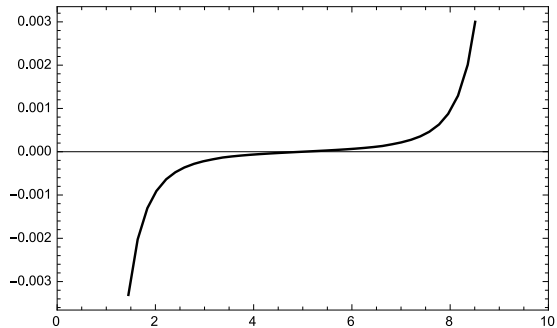}
\caption{Change in the Casimir energy for $\epsilon=0.3$ }
\label{mixcas}
\end{figure}

With this, we have that the instantaneous rate of change in the Casimir energy given in Figure \ref{mixcas}. 
As opposed to the previous analysis, here we have that the change in the Casimir energy changes sign at the middle of the interval.  This change depends on the proximity to the edges, being the closest the dominant one. 

In this case, we have that the behavior of the change in the Casimir energy is that it is odd with respect to the midpoint of the interval, when regarded as a function of $t$. As opposed to the previous scenario, we have that the change in the Casimir energy changes sign at the middle of the interval, having an apparent point of inflection here. 

\section{Conclusions}
In this work we found the impact that can have a localized perturbation on the Casimir energy. This highly depends on the profile function and the shape of the perturbation function, as well as on the proportion between the length domain interval and the size of the profile function. 

Through the numerical analysis, we found that the sign of the change in the Casimir energy does not depend entirely on the shape of the profile function function, but on the proportion between the size of the profile function and the length of the domain interval, having a great impact the distance to the edges of the surface to the localized perturbation.

This result confirms that the Casimir energy mainly depends on the local geometry, rather than global properties. An example of this is the case where the length of the interval is considerably bigger than the size of the profile function, where the change in the Casimir energy behaves like the one of an infinite cylinder.


\begin{thebibliography}{}
\bibitem{Abalo10} Abalo, E. K. and Milton,K. A. ,\emph{Casimir energies of cylinders: Universal function}, Phys. Rev. D 82, 125007 \textbf{(2010)}

\bibitem{Abalo12} Abalo, E. K.  \emph{et al}, \emph{Scalar Casimir energies of tetrahedra and prisms}, J. Phys. A: Math. Theor. 45 425401 \textbf{(2012)}


\bibitem{Bush92} Bush, A. W. \emph{Perturbation Methods for Engineers and Scientists}, CRC Press, Jan 21, \textbf{(1992)}


\bibitem{Casimir48} Casimir, H. B. G. . \emph{On the attraction between two perfectly conducting plates}. Proc. Kon. Nederland. Akad. Wetensch. B51: 793 \textbf{(1948)}

\bibitem{CasimirP48} Casimir, H. B. G. and Polder, D. \emph{The Influence of Retardation on the London-van der Waals Forces}, Phys. Rev. 73, 360–372 \textbf{(1948)}

\bibitem{Dowker76}Dowker, J.S. and Critchley, R., \emph{Effective Lagrangian and energy-momentum tensor in de Sitter space}, Phys. Rev.D 13, 3224 \textbf{(1976)}


\bibitem{Euler44} Euler, L., \emph{Variae observationes circa series infnitas}, Commentarii academiae scientiarum Petropolitanae 9,  160-188 \textbf{(1744)}

\bibitem{Euler60} Euler, L., \emph{De seriebus divergentibus}, Novi Commentarii academiae scientiarum Petropolitanae 5, 205-237 \textbf{(1760)}


\bibitem{Gosdzinsky98} Gosdzinsky P., Romeo A., \emph{Energy of the vacuum with a perfectly conducting and infinite cylindrical surface}, Physics Letters B, Volume 441, Issues 1–4, Pages 265-274 \textbf{(1998)}


\bibitem{Hardy16}Hardy, G.H. and Littlewood, J.E., \emph{Contributions to the Theory of the Riemann Zeta-Function and the Theory of the Distribution of Primes}, Acta Mathematica, 41 pp. 119–196 \textbf{(1916)}

\bibitem{Hawking77}Hawking, S. W., \emph{Zeta function regularization of path integrals in curved spacetime}, Communications in Mathematical Physics 55 (2): 133–148 \textbf{(1977)}

\bibitem{Jeffres12} Jeffres, Thalia, D., Kirsten,K., and Lu, T., \emph{Zeta Function on Surfaces of Revolution}, arXiv:1211.4043v1 \textbf{(2012)}

\bibitem{Kenneth02} Kenneth, O., Klich, I., Mann, A., and Revzen, M. ,\emph{Repulsive Casimir Forces}, Phys. Rev. Lett. 89, 033001 \textbf{(2002)}

\bibitem{Kimball12} Milton, K. A. \emph{et al} \emph{Repulsive Casimir and Casimir–Polder forces}, J. Phys. A: Math. Theor. 45 374006 \textbf{(2012)}


\bibitem{Lamoreaux97} Lamoreaux, S. K., \emph{Demonstration of the Casimir Force in the 0.6 to 6 $\mu$m Range}, Phys. Rev. Lett. 78, 5–8 \textbf{(1997)}

\bibitem{Riemann92} Riemann, B.. \emph{\"Uber die Anzahl der Primzahlen unter einer gegebenen Gr\"osse}, Monatsberichte der Berliner Akademie \textbf{(1859)}

\bibitem{Shivamoggi03}Shivamoggi, Bhimsen, \emph{Perturbation Methods for Differential Equations},  XIV, Birkhäuser Basel \textbf{(2003)}



\end{thebibliography}
\end{document}